\documentclass[epj]{svjour}

\usepackage{graphicx}
\usepackage{fancyhdr}
\usepackage{amsmath}
\usepackage{amssymb}
\usepackage{bbm}
\def\makeheadbox{{
 }}
\def\mytitle{My title} 
\def\myauthors{My name}  
\def\mytype{My type of session}
\def\mysession{My session}

\def\hbar{\hspace{0pt}\raisebox{1pt}{$-$} \hspace{-7pt} h}

\newcommand{\be}{\begin{equation}}
\newcommand{\ee}{\end{equation}}
\newcommand{\bd}{\begin{displaymath}}
\newcommand{\ed}{\end{displaymath}}
\newcommand{\bea}{\begin{eqnarray}}
\newcommand{\eea}{\end{eqnarray}}
\newcommand{\nn}{\nonumber}

\def\mytitle{SUSY flavour \& CP problem - SUGRA versus SU(3)} 
\def\myauthors{Michal Malinsk\'y}    
\def\mytype{Contributed Talk}    
\def\mysession{Flavor Physics}

\begin{document}

\title{Tackling the SUSY flavour \!\&\!  CP problem \!-\! SUGRA vs. SU(3)}
\author{Michal Malinsk\'y\inst{1}
\thanks{\emph{Email:} malinsky@phys.soton.ac.uk}%
}                     
\institute{School
of Physics and Astronomy, University of Southampton, SO17 1BJ
Southampton, United Kingdom}
%
\date{}
\abstract{
We comment on the power of the 'standard' solutions to the SUSY flavour and CP problem based on supergravity and its derivates like mSUGRA in comparison to the flavour symmetry approach. It is argued that flavour symmetries, and $SU(3)_f$ in particular, can not only mimic the situation in supergravity frameworks in this respect, but sometimes do even better providing at the same time a further link between the soft and Yukawa sectors, that could be testable at future experimental facilities.   
} 
\maketitle
%
\section{Introduction}
Though incredibly successfull in many aspects, the Standard model of elementary particle interactions (SM) can not be regarded to as a final theory of nature. One of its deepest mysteries is the origin of fermion masses and mixings, in particular the enormous hierarchy between the unprecedentally light neutrinos \cite{Maltoni:2004ei,Strumia:2006db} at one side and the top-quark mass in the vicinity of the electroweak scale on the other. Another facet of this so called `SM flavour problem' is the lack of understanding the origin of the peculiar mixing paterns in charged current interactions governed in the quark sector by the Cabibbo-Kobayashi-Maskawa (CKM) matrix not far from unity, while the corresponding mixing in the lepton sector features a nearly tri-bimaximal patern \cite{Harrison:2002er} with two large angles entering the corresponding MNS \cite{Maki:1962mu} lepton mixing matrix.

While the overall smallness of the neutrino masses can be accounted for by invoking the SM-allowed dimension 5 effective operators of the type $L H L H/{\Lambda}$  \cite{Weinberg:1980bf} finding naturally its dynamical justification in a class of seesaw models \cite{Minkowski:1977sc} with $\Lambda$  being a new physics scale at around $10^{14}$~GeV  (remarkably close to the one suggested by the gauge coupling unification), the understanding of the striking difference between the quark and lepton spectra and mixing patterns require a more detailed quantitative approach.
One of such, comming under the names of Froggatt and Nielsen \cite{Froggatt:1978nt}, offers a simple  interpretation of the hierarchies among the SM fermion families as a consequence of brekdown of an extra `flavour' symmetry. This is usually triggered by flavour-nontrivial VEVs of  additional Higgs (flavon) scalars and subsequently transmitted to the matter sector via flavon-matter interactions respecting the flavour structure of this extra symmetry. The Yukawa couplings are then generated at non-renormalizable level and their hierarchy emerges from the powers of ratios of the flavon VEVs over the masses of the underlying `messenger' fields giving rise to the relevant effective operators.

However, it is actually quite difficult to construct simple and potentially realistic models of flavour along these lines. In order to account for all the quantitative subtleties in the quark and lepton spectra and mixings one is necessarily lead to rather complicated constructs featuring multiple flavon fields triggering the various flavour symmetry breaking steps. This, however, impacts the predictive power of such models considerably and one has to compromise on either the predictivity or the precision side. 
Unfortunately, there is only few indications from the physical data what the underlying flavour symmetry (and its breaking pattern) could be. Perhaps the most solid piece of information available at the SM level comes from the tri-bimaximal lepton mixing pattern that indicates a non-abelian nature of the flavour group\footnote{The typical $\tfrac{1}{\sqrt{2}}$ and $\sqrt{\tfrac{2}{3}}$ entries in the tri-bimaximal lepton mixing matrix resemble the Clebsch-Gordon coefficients associated to the irreps of non-abelian Lie-algebras.}. Moreover, in the context of the so called sequentially dominated seesaw \cite{King:1999cm,King:1999mb} the strong correlations between the elements of a typical neutrino Yukawa matrix across all three families suggests for a 'maximal' (i.e. universal) flavour symmetry to be restored at a high scale, like e.g. $SU(3)$, $SO(3)$ or some of their discrete subgroups.   

The high level of ambiguity is clearly mirrored in the proliferation of studies based on different assumptions on the actual shape of the flavour symmetry breaking pattern. An iterested reader can find a representative sample in e.g. \cite{Bando:2001bj,King:2000ge,Ling:2002nj,Kane:2005va,Grimus:2002zh,Blazek:1999hz,Raby:2003ay,King:2003rf,Barbieri:1999km,Antusch:2004xd,King:2006me,Hagedorn:2006ir,Grimus:2004rj,Dermisek:2005ij,Dermisek:2006dc,Babu:2002dz,King:2006np,Altarelli:2005yx,Altarelli:2006kf} and references therein. Thus, in order to tell a particular model from the others, extra inputs (apart from the limited number of flavour parameters of the SM) are certainly needed.

If the low-scale supersymmetry (SUSY) is invoked, there is a variety of extra constraints on the flavour structure of a given model associated to the SUSY flavour and CP problem. Even in the simplest potentially realistic SUSY scenario, the minimal supersymmetric extension of the Standard model (MSSM), there are soft-SUSY breaking terms in the lagrangian that constitute a new source of flavour and CP physics beyond the SM framework. Denoting the matter piece of the MSSM superpotential by
\bea
\label{Wmatter}
W_{m} & = & \varepsilon_{\alpha\beta} \left[
\hat{H}_u^\alpha \hat{Q}^{\beta i} Y^{u}_{ij} \hat{u}^{cj} + \hat{H}_d^\alpha
\hat{Q}^{\beta i} Y^{d}_{ij} \hat{d}^{cj} + \right.
 \\
&+& \left.
\hat{H}_u^\alpha
\hat{L}^{\beta i} Y^{\nu}_{ij} \hat{N}^{cj}+ \hat{H}_d^\alpha
\hat{L}^{\beta i} Y^{e}_{ij} \hat{e}^{cj} \right]+\hat N^{ci}M^{\nu}_{ij}\hat N^{cj} \nn
\eea 
(with hats for superfields), the soft-SUSY breaking lagrangian driving the scalar (superpartner) sector reads
\begin{eqnarray}
\label{Lsoft}
{\cal L}_\mathrm{soft} &\ni& \varepsilon_{\alpha\beta} \left[ 
{H}_u^\alpha \tilde{Q}^{\beta i}  A^{u}_{ij} \tilde{u}^{cj}
+ {H}_d^\alpha \tilde{Q}^{\beta i}  A^{d}_{ij} \tilde{d}^{cj}\right. \\
& + & \left.
 {H}_u^\alpha \tilde{L}^{\beta i}  A^{\nu}_{ij} \tilde{N}^{cj}
+ {H}_d^\alpha \tilde{L}^{\beta i}  A^{e}_{ij} \tilde{e}^{cj} 
+ \text{h.c.}\right]\nonumber \\
& + & 
\tilde{Q}_{i\alpha}^{*} ( m_{Q}^2)^{i}_{j} \tilde{Q}^{\alpha j}
+ \tilde{u}^{c *}_i ( m_{u^{c}}^2)_{j}^{i} \tilde{u}^{cj}
+ \tilde{d}^{c *}_i ( m_{d^{c}}^2)_{j}^{i} \tilde{d}^{cj}\nn \\
&+ &\tilde{L}_{i\alpha}^{*} ( m_{L}^2)^{i}_{j} \tilde{L}^{\alpha j}
+ \tilde{e}_i^{c *} ( m_{e^{c}}^2)_{j}^{i} \tilde{e}^{cj}
 +  \tilde{N}_i^{c *} ( m_{\nu^{c}}^2)_{j}^{i} \tilde{N}^{cj}\nn
\end{eqnarray}
The strong suppression of flavour-changing neutral current (FCNC) interactions and the striking compatibility of the whole plethora of the CP violating phenomena with the simplest CKM hypothesis puts stringent constraints on the alignment of the Yukawa structures in (\ref{Wmatter}) to the corresponding trilinears ($A$-terms) in (\ref{Lsoft}). Simultaneously, the soft mass-terms in (\ref{Lsoft}) generating the ($6\times 6$) scalar masses 
\bea
M^{2}_{\tilde{f}}& \equiv & \begin{pmatrix}
{m^2_{\tilde{f}}}_{LL} & {m^2_{\tilde{f}}}_{LR} \\
{m^2_{\tilde{f}}}_{RL} & {m^2_{\tilde{f}}}_{RR} 
\end{pmatrix}\nn \quad \mathrm{provided}
\\
m^{2}_{\tilde{f}LL,RR}& = &{m}^{2}_{f,f^{c}}+ (Y^{f} Y^{f\dagger},  Y^{f\dagger} Y^{f})v_{u,d}^{2}+ D^{f}_{LL,RR}\nn \\
m^{2}_{\tilde{f}LR} & = & A^{f}v_{u,d}-\mu Y^{f}v_{d,u}, \quad m^{2}_{\tilde{f}RL}=(m^{2}_{\tilde{f}LR})^{\dagger}\nn
 \eea
($D^{f}_{XY}$ with $X,Y\in \{L,R\}$ denoting the chiralities stands for the various D-terms, $\mu$ is the $\mu$-term) should receive an approximately flavour diagonal form. A convenient parametrization is obtained upon adopting the so-called Super-CKM (SCKM) basis \cite{Dugan:1984qf} where all the Yukawa matrices are diagonal ($\tilde{Y}^{f}\equiv U_{L} Y^{f}U_{R}^{f\dagger}$) and the extra SUSY flavour  and CP effects can be encoded in the flavour off-diagonalities and phases of entries of  $\tilde{M}^{2}_{\tilde{f}}\equiv (U_{L}^{f}, U_{R}^{f})M^{2}_{\tilde{f}}(U_{L}^{f}, U_{R}^{f})^{\dagger}$. Since so far there has been no such beyond-SM flavour/CP effect  observed, stringent upper bounds are imposed on the `normalized' flavour and CP violating mass insertions $(\delta^{f}_{XY})_{ij}\equiv (\tilde{m}^{2}_{\tilde{f}XY})_{ij}/\langle\tilde{m}^{2}_{\tilde{f}XY}\rangle$. The issue then is to understand the several orders of magnitude suppression of the physical $\delta^{f}_{XY}$'s with respect to their natural ${\cal O}(1)$ values. 

The traditional way this problem is addressed rely on extra assumptions about the origin of flavour violation in the soft sector. For instance, in the minimal flavour violation (MFV) scenario, all the flavour violation is associated to the Yukawa sector \cite{Ciuchini:1998xy}. Even more ambitiously, one often just puts by hand all the trilinears to be proportional to the associated Yukawas and the soft masses $m_{f}^{2}\propto \mathbbm{1}$  (at the SUSY breaking scale) under the name of minimal supergravity (mSUGRA) ansatz, that can be further justified and augmented in models where the underlying dynamics responsible for transmitting the SUSY breakdown from a hidden sector to matter is flavour blind, e.g. various gauge-mediation scenarios, gravity mediation in the supergravity framework etc.     

For simplicity reasons, in this short paper we shall focus on the SUGRA \cite{Dugan:1984qf,Hall:1983iz} solution (and partially also on its derivates like mSUGRA). It is not only concise, but also (given the K\"ahler potential) brings in the virtue of straightforward soft-sector calculability. If, for instance, the K\"ahler potential is of the ``sequesterd'' form , i.e.
$
K=\tilde{K}_{\overline{a}b}f^{\overline{a}*}f^{b}+K_{hid}
$ 
(where $\tilde{K}_{\overline{a}b}$ denotes the matter sector K\"ahler metric) and only the `hidden' piece $K_{hid}$ is sensitive to the SUSY-breaking driven by a hidden sector field $X$, and if also the matter superpotential contains no direct couplings to $X$, then the SUGRA-induced effective soft-term formulae   
\bea
m^{2}_{\tilde{f}} & = & m_{3/2}^{2}\tilde{K}_{\overline{a}b}\label{SUGRAsoftterms0} \\
&- &\!\! \sum_{S,S^{'}}F_{\overline{S}'}\!
\left[\partial_{\overline{S}'} \partial_{S}{\tilde{K}}_{\overline{a}b}-\partial_{\overline{S}'}{\tilde{K}}_{\overline{a}c}
({\tilde{K}}^{-1})_{c\overline{d}}
\partial_{S}{\tilde{K}}_{\overline{d}b}
\right]\!
F_{S}\nn \\
{A}_{abc}& \propto & 
\sum_S F_S\left\{
\tfrac{1}{M_{Pl}^{2}}(\partial_{S}K_{hid.})Y_{abc}+\partial_{S}Y_{abc}\right.  \label{SUGRAsoftterms}\\
&-& \left. \left[
(\tilde{K}^{-1})_{d\overline{e}}\partial_{S}\tilde{K}_{\overline{e}a}Y_{dbc}
+cycl.\right] \nonumber
\right\}
\eea
lead exactly to the desired (SUSY-breaking scale) pattern $A^{f}\propto Y^{f}$ and\footnote{Performing the canonical normalization of the matter sector kinetic terms, one gets $m^{2}_{{f}}\propto \mathbbm{1}$ along the mSUGRA ansatz.}   $m^{2}_{{f}}\propto \tilde{K}$. Note that the summation in (\ref{SUGRAsoftterms0}) and (\ref{SUGRAsoftterms}) is taken over all the fields $S$, $S'$ developing an $F$-term; for the current example $S, S'=X$.

Let us see what happens when one attempts to marry the flavour symmetries (providing a handle on the Yukawa patterns) with the SUGRA solution to the SUSY flavour and CP problem. 
First, the K\"ahler potential is not imune to any source of flavour breaking (e.g. the flavon VEVs) and the canonical normalization procedure tends to destabilize any (just by hand imposed) proportionality relations \`a la mSUGRA etc. 

Can then a full-fledged SUGRA with sequestered K\"ahler and superpotential help to keep these corrections under control and restore the high-scale alignment on dynamical grounds? Remarkably enough, the answer is no \cite{inpreparation} due to the observation \cite{Ross:2002mr,Ross:2004qn} that every superfield whose scalar component receives a VEV  tends to develop a SUGRA $F$-term. Such an $F$-term subsequently enters the effective soft-sector formulae (\ref{SUGRAsoftterms0}), (\ref{SUGRAsoftterms}) with the potential to lift again the dynamical (sequestered) supergravity alignment. 

\section{Irreducible flavon $F$-terms in SUGRA }
Recalling the generic prescription for a supergravity $F$-term: 
\be\label{Ftermshape}
F_{S}=- (K^{-1})_{S\overline{S}'}\left(m_{3/2}{K}_{\overline{S}'}-e^{K/{2M_{Pl}^{2}}}W^{*}_{\overline{S}'}\right)
\ee
where $m_{3/2}$ is the gravitino mass (defined as $m_{3/2}^{2}=e^{\langle  K/{M_{PL}^{2}}\rangle}\langle{|W|^{2}}/{M_{Pl}^{4}}\rangle $), one can see that for any scalar receiving a VEV the first term in the parenthesis in (\ref{Ftermshape}) is nonzero\footnote{Recall that no symmetry can prevent K\"ahler from containing terms like $S S^{*}$} and (barring in mind the further complication arising from a would-be non-canonical shape of the K\"ahler and possible cancellations between the two terms above) leads to an irreducible contribution to a flavon $F$-term of the form   
$\langle F_{\phi}\rangle \propto m_{3/2}\langle\phi\rangle$.

It was argued \cite{Ross:2002mr,Ross:2004qn} that these terms could easily compete with the leading order hidden sector contributions to the effective $A$-terms in formula (\ref{SUGRAsoftterms}). Indeed, 
the shape of the second term in (\ref{SUGRAsoftterms}) is such that the extra flavon insertion brought in by the flavon $F$-term is exactly compensated in the derivative of the Yukawa, thus keeping the overall magnitude of  the leading $F_{\phi}\partial_{\phi}Y$ term intact. This can have implications for the $A$-term driven flavour and CP violating parameters, see e.g. \cite{Ross:2002mr,Ross:2004qn}.  
On the other hand, due to the presence of the universal leading term in relation (\ref{SUGRAsoftterms0}), the irreducible flavon contribution enters the effective soft-masses only at the subleading level and thus is usually ignored.

However, even such a subleading contribution can render the supergravity framework with flavour symmetries essentially indistinguishable from the effective K\"ahler-potential-driven flavour violation along the lines of mSUGRA \cite{inpreparation}. Moreover, if the flavour symmetry happens to mimic also the leading-order flavour-diagonality of the effective soft-mass term expansion (\ref{SUGRAsoftterms}), as for instance $SU(3)_f$ does, there could be even no reason to employ SUGRA as an aid with the SUSY flavour and CP problem - the flavour violation generated in an effective flavour-symmetry governed framework would just resemble the irreducible flavon $F$-term-induced flavour violation in the full-fledged supergravity \cite{inpreparation}.

\section{Effective $SU(3)$ flavour symmetry }
In what follows, we shall explore the situation in a particular flavour model \cite{deMedeirosVarzielas:2005ax,deMedeirosVarzielas:2005qg} where the $SU(3)_f$ flavour symmetry is promoted to the effective soft-SUSY breaking sector. We shall assume the most general $SU(3)_f$-compatible soft-sector operator expansions \cite{Antusch:2007re} without any reference to a particular SUSY-breaking dynamics. As we shall see, a pure flavour symmetry can alleviate the flavour and CP issue even better than mSUGRA \cite{Antusch:2007re}  and its marriage with the full-fledged supergravity need not bring any further relief either \cite{inpreparation}. This, however, makes the simplest supergravity  flavour models potentially testable in the near future.

The model under consideration employs a relatively simple pattern of $SU(3)_{f}$-antitriplet flavon fields $\phi_{123}$, $\phi_{23}$ and $\phi_{3}$ (together with their hermitean conjugate triplets protecting the $D$-flattness downto the electroweak scale) developing their VEVs along the different (anti)triplet directions:
\begin{eqnarray}\label{VEVsinthegoodbasis}
& \langle\phi_{123}\rangle = (
1,
e^{i\phi_{1}},
e^{i\phi_{2}})^{T}
u_{1}\; ,
\langle\phi_{23}\rangle = (
0,
1,
e^{i\phi_{3}})^{T}u_{2}\; , & \nn\\
&
\langle\phi_{3}\rangle^{u,d} = (
0, 
0,
1)^{T}u_{3}^{u,d}\nn.
&
\end{eqnarray}
The effective Yukawa superpotential is driven by extra symmetries, c.f. \cite{deMedeirosVarzielas:2005ax,deMedeirosVarzielas:2005qg,Antusch:2007re} into the following form:
\bea
\label{TBYukawa} 
W_{Y}&=&
f^{i}f^{cj}
\left[y_1^f(\phi_{123})_{i}(\phi_{23})_{j}+y_2^f(\phi_{23})_{i}(\phi_{123})_{j}\right.\nn\\
&+& \left.y_3^f(\phi_{3})_{i}(\phi_{3})_{j}+{y_{4}^f}(\phi_{23})_{i}(\phi_{23})_{j}\right]\tfrac{H}{M_{f}^{2}}+\ldots\;\;\;\;
\eea
(with $M_{f}$ denoting the messenger masses) which leads to a realistic shape of all the Yukawa matrices provided 
${u_{2}}/{M_{u,Q,\nu,L}} =\varepsilon$,  
${u_{2}}/{M_{d,e}} = \bar\varepsilon$, 
${u_{1}}/{M_{u,Q,\nu,L}} = \varepsilon^2$ and 
${u_{1}}/{M_d,e} = \bar\varepsilon^2 $,
where the two expansion parameters obey $\varepsilon \approx 0.05, \bar\varepsilon\approx 0.1\sim 0.15$, see e.g. \cite{Roberts:2001zy}. 
Note that in order to reconstruct the tri-bimaximal mixing pattern in the lepton sector, it is convenient to make the last term in eq. (\ref{TBYukawa}) vanish for neutrinos \cite{King:1999cm}, that could be achieved by promoting it to a higher order operator: 
\be\label{Sigma}
\frac{y_{4}^f}{M_{f}^{2}}(\phi_{23})_{i}(\phi_{23})_{j}H=
\frac{y_{\Sigma}}{M_{f}^{2}M_{\Sigma}}(\phi_{23})_{i}(\phi_{23})_{j}\Sigma H
\ee
where $\Sigma$ is a flavour-blind Higgs with zero Clebsch-Gordon coefficitent in the neutrino direction. 

Promoting the $SU(3)_{f}$ flavour symmetry into the effective soft-SUSY-breaking sector, one can write at the leading order
\bea
\label{Aexpansion}
\tfrac{\hat{A}^{f}_{ij}}{A_{0}} & =  & a^{f}_{1}\tfrac{\langle\phi_{123}\rangle_{i} \langle\phi_{23}\rangle_{j}}{{M_f}^2}+a^{f}_{2}\tfrac{\langle\phi_{23}\rangle_{i} \langle\phi_{123}\rangle_{j}}{{M_f}^2}\nn \\
&+& a^{f}_{3} \tfrac{\langle\phi_{3}\rangle_{i} \langle\phi_{3}\rangle_{j}}{{M_f}^2}+a_{\Sigma}^{f}\tfrac{\langle\phi_{23}\rangle_{i} \langle\phi_{23}\rangle_{j}\langle\Sigma\rangle}{{M_f}^2M_{\Sigma}}+\ldots \nn
\eea
for the $A$-terms, in full analogy with the Yukawa sector expansion stemming from (\ref{TBYukawa}) and (\ref{Sigma}), while the soft masses and the K\"ahler potential obey (notice the leading order universality due to the $SU(3)_{f}$ symmetry):
\bea
(\hat m^{2}_{f,f^{c}})_{ij} &=& 
m_{0}^{2}\left( b_{0}^{f,f^{c}}\delta_{ij} + 
b^{f,f^{c}}_{1}\tfrac{\langle\phi_{123}\rangle_{j}\langle\phi_{123}^{*}\rangle_{i}}{{M_{f}}^2}\right.\nn \\
& + &\left.
b^{f,f^{c}}_{2}\tfrac{\langle\phi_{23}\rangle_{j}\langle\phi_{23}^{*}\rangle_{i}}{{M_{f}}^2}+ 
b^{f,f^{c}}_{3}\tfrac{\langle\phi_{3}\rangle_{j}\langle\phi_{3}^{*}\rangle_{i}}{{M_{f}}^2} 
\right)+\ldots\nn\\
(\tilde K_{f,f^{c}})_{ij} &=& 
k_{0}^{f,f^{c}}\delta_{ij} + 
k^{f,f^{c}}_{1}\tfrac{\langle\phi_{123}\rangle_{j}\langle\phi_{123}^{*}\rangle_{i}}{{M_{f}}^2}\nn \\
& + &\! 
k^{f,f^{c}}_{2}\tfrac{\langle\phi_{23}\rangle_{j}\langle\phi_{23}^{*}\rangle_{i}}{{M_{f}}^2}+ 
k^{f,f^{c}}_{3}\tfrac{\langle\phi_{3}\rangle_{j}\langle\phi_{3}^{*}\rangle_{i}}{{M_{f}}^2} 
+\ldots\nn
\eea
The hats in $\hat A$ and $\hat m^{2}$ denote a quantity before the canonical normalization conditions are imposed. The naturalness then requires that all the $y_{i}$, $a_{i}$, $b_{i}$ and $k_{i}$  coefficients drop into the ${\cal O}(1)$ domain. Along the lines of \cite{deMedeirosVarzielas:2005ax,deMedeirosVarzielas:2005qg,Antusch:2007re} we shall further assume that CP is only spontaneously broken in the flavon sector. A complete analysis of these structures is beyond the scope of this work and can be found 
in \cite{Antusch:2007re}.

In what remains, let us focus only on the current model predictions for the electric dipole moments of electron and neutron. Being strongly suppressed in the SM, these quantities are indeed  very sensitive to the effective soft-SUSY-breaking sector flavour structure and can be measured rather accurately. The current bounds $d_{e}<6.3\times 10^{-26} e\, \mathrm{cm}$, $d_{n}<4.3\times 10^{-27} e\, \mathrm{cm}$ correspond to the constraints on the relevant $\delta$-parameters (for $\langle m_{\tilde q}\rangle\sim 500$ GeV and $\langle m_{\tilde l}\rangle\sim 100$ GeV) 
of the form $|\mathrm{Im}(\delta_{11}^{u,d})_{LR}|\lesssim 10^{-6}$,  $|\mathrm{Im}(\delta_{11}^{l})_{LR}|\lesssim 10^{-7}$ \cite{Abel:2001vy}.
Defering the computation details into \cite{Antusch:2007re}, one receives approximately (forgetting about the subleading effects due to canonical normalization)
\bea & & 
|\mathrm{Im}(\delta_{11}^{u})_{LR}|\sim 10^{-7}\,\tfrac{A_{0}a^{u}_{\Sigma}}{\langle\tilde m_{u}\rangle}\left|\tfrac{a^{u}_{1}+a^{u}_{2}}{a^{u}_{\Sigma}}-\tfrac{y^{u}_{1}+y^{u}_{2}}{y^{u}_{\Sigma}}\right|\sin\phi_{1}\;\;\;\;\;\;\; \label{EDMs}\\
& & |\mathrm{Im}(\delta_{11}^{d})_{LR}|\sim 5.10^{-6}\,\tfrac{A_{0}a^{d}_{\Sigma}}{\langle\tilde m_{d}\rangle}\left|\tfrac{a^{d}_{1}+a^{d}_{2}}{a^{d}_{\Sigma}}-\tfrac{y^{d}_{1}+y^{d}_{2}}{y^{d}_{\Sigma}}\right|\sin\phi_{1}\nn \\
& & |\mathrm{Im}(\delta_{11}^{l})_{LR}|\sim 5.10^{-6}\,\tfrac{A_{0}a^{l}_{\Sigma}}{\langle\tilde m_{l}\rangle}\left|\tfrac{a^{l}_{1}+a^{l}_{2}}{a^{l}_{\Sigma}}-\tfrac{y^{l}_{1}+y^{l}_{2}}{y^{l}_{\Sigma}}\right|\sin\phi_{1}\nn
\eea
where $\phi_{1}$ is the CP phase associated to the VEV of $\phi_{123}$. Unlike in mSUGRA, where the leading contributions emerge at the {\it quartic} level in the small parameters ${\varepsilon}$, $\overline{\varepsilon}$, the overall magnitude of the numerical factors above come from the {\it fifth} power \cite{Antusch:2007re} of $ \varepsilon$ (for up-sector) and $\overline{\varepsilon}$ (for down quarks and charged leptons, further suppressed by $\tan\beta$ in the denominator)
which (taking into account the running effects) is compatible with the low-energy experimental bounds.

Let us emphasize that these results are driven by the flavour symmetry only. All that SUGRA can further do is constraining the alignment\footnote{Note that if $A^{f}\propto Y^{f}$, the leading order contributions to EDMs given above vanish as expected.} of the effective couplings $a_{i}$ and $y_{i}$. However, applying formula (\ref{SUGRAsoftterms}) for the irreducible flavon $F$-terms, one obtains different proportionality factors for $a_{1,2}=k_{12} y_{1,2}$ and $a_{\Sigma}=k_{\Sigma} y_{\Sigma}$ with $k_{\Sigma}=\tfrac{3}{2}k_{12}$ and thus SUGRA does not make any of the parentheses in (\ref{EDMs}) vanish! The reason is that $(\delta_{LR}^{f})_{11}$ in the $SU(3)_{f}$ model under consideration come from an interplay between effective operators of different dimensionalities that in formula (\ref{SUGRAsoftterms}) cause different scaling behavior of the associated supergravity driven $A$-term coefficients.

\section{Conclusions}
Thus, as far as the flavour and CP violating quantities coming from interplay of operators of different dimension in the effective expansion are taken into account, an $SU(3)_f$ flavour symmetry could be at least as good solution to the SUSY flavour and CP problem as  the full-fledged supergravity. The reason is the presence of irreducible flavour and CP-violating effective soft-sector terms associated to the inherent flavon-induced $F$-terms in supergravity models of flavour.
\subsection*{Acknowledgements}
The author is very grateful to Steve King, Stefan Antusch and Graham Ross for invaluable discussions. The work was particularly supported by PPARC Rolling Grant PPA/G/S/2003/00096. 

\end{document}